\documentclass{article}
\usepackage{amsmath}
\usepackage{amssymb}
\usepackage{amsthm}
\usepackage{graphicx}
\usepackage{psfrag}
\usepackage[hang, nooneline]{subfigure}
\usepackage{color}

\author{Natascha Riahi \footnote{e-mail address: natascha.riahi@gmx.at}
\\ University of Vienna, Faculty of Physics, Gravitational Physics
\\ Boltzmanng. 5, 
1090 Vienna, Austria}
\date{}
\title{Analysis of wavepacket tunneling with the method of Laplace transformation}

\begin{document}
\maketitle
\begin{abstract}
 We use the method of Laplace transformation to determine the dynamics of a wave
packet that passes a barrier by tunneling. We investigate the transmitted wave packet
and find that it can be resolved into a sequence of subsequent wave packages. This result
sheds new light on the Hartman effect for the tunneling time and gives a possible explanation for an 
experimental result obtained by Spielmann et. al.
\end{abstract}

\section{Introduction}

There are several definitions
 of tunneling times (\cite{Hauge},\cite{Landauer},\cite{Winful},\cite{Olkhovsky},\cite{Razavy},\cite{Galapon},\cite{Landsmann},\cite{Landsmann1})
 and the discussion about their meaning is still 
ongoing
(\cite{Winful},\cite{Olkhovsky},\cite{Wang},\cite{Cheng}). Recently, 
experiments with atoms stimulated by ultrashort, 
infrared laser pulses, the so-called attoclock experiments,
reinforced the interest in the prediction of tunneling times (\cite{Pfeiffer},\cite{Eckle},\cite{Landsmann2}).

In this article we  investigate the dynamics of a wavepacket that tunnels through a barrier.
The tunneling time we determine is the
 so-called group delay or phase time (\cite{Hartmann},\cite{Winful}). It can be characterized as the time interval
 between the moment the peak of a freely evolving incoming  wave packet would reach the barrier and the arrival of the peak of the transmitted
 wave packet at the end of the barrier.
 One of the oldest results for this tunneling time was obtained by Mac Coll 
(\cite{MacColl}) in 1932 by the application of the stationary phase method. He concluded
that there was ''no appreciable delay in the transmission of the packet through the barrier''. These calculations were later 
refined by Hartmann (\cite{Hartmann},\cite{Winful}), 
who found a finite delay time. For thicker barriers this delay
time becomes independent of the thickness of the barrier and tends to a fixed value
which is known as the Hartmann effect. This saturation  was also found
for other definitions of tunneling time (\cite{Olkhovsky}), whereas in the framework of fractional quantum mechanics a decreasing of the 
tunneling time with barrier width was obtained(\cite{Hasan}).  A formal analogy between the Schr\"odinger and the 
Helmholtz equation (see for instance \cite{Winful})made it possible to test predictions for the tunneling time with optical experiments. 
The results confirmed the saturation of the tunneling time (\cite{Steinberg},\cite{Spielmann}, \cite{Longhi},see \cite{Winful} for more references ) 
though a more detailed discrepancy between experiment and calculations remained open in \cite{Spielmann}.

 Since the Hartmann effect seems to threaten Einstein causality, because superluminal velocities can be infered from transmission times saturating with barrier width ,  an objection was made that only 
 the modes above the barrier energy may contribute to the transmitted wave packet
 (\cite{Steinberg}, \cite{Chiao}). But it was argued in (\cite{Winful1})with reference to numerical calculations and experimental results that this effect can not be used as an explanation of the barrier tunneling phenomenon  (see also \cite{Winful} for a more detailed discussion and further references).

The results of the attoclock experiments with strong laser fields that lower the Coulomb potential  can be best explained by using a 
tunneling time probability amplitude constructed with Feynnman path integrals (\cite{Pfeiffer}, \cite{Landsmann2}), whereas several other  definitions of tunneling times do 
 not agree with the data.
 
 At present there exists no unified  formalism for the calculation of tunneling times for different experimental settings. 
 In the case of wavepacket tunneling through static barriers the phase time, keeping track of the peak is the best candidate for the tunneling time, 
 and in good agreement with the experimental results(\cite{Steinberg},\cite{Spielmann}, \cite{Longhi}).

The phase time is in general obtained by representing the solution of the 
 time-dependent Schr\"odinger equation as the integral over stationary solutions for different energies. 
Under the assumption that the
energy distribution of the initial wave packet is sufficiently peaked the application of the 
stationary phase
method yields a time for the arrival of the peak of the wavepacket at the end of the barrier. 

In this article we use the method of Laplace transformation
\cite{Methode} to determine the wavepacket dynamics.Instead of immediately applying an approximation for oscillatory integrals, we first obtain exact solutions for the wavefunction at each side and inside  the barrier, which we then simplify assuming that the initial wavepacket is sufficiently 
 peaked in the momentum 
representation. Since we are free to choose explicitely the initial wavepacket
in position space (which we do not have to construct as superposition of eigenstates), it is
technically much easier to start with a wavepacket located exclusively at the left side of the barrier. 
Furthermore we give estimates for the consequences of the applied approximations and derive a consistency condition in the case of a special class of initial wavepackets.

We find that what comes out of the barrier is not a 
single
wave packet, but infinitely many, one after the other. The time each of these  wavepackets needs 
for
the tunneling 
is completely independent of the thickness of the barrier. But for thicker barriers, the later
wavepackets are much more attenuated and so there remains a single transit time that equals the result Hartmann obtained as 
the upper limit for the tunneling time. However in our approach the tunneling decreases before it reaches the limiting value
for thick barriers. This is in accordance with the results of the optical experiments performed by
Spielmann et al. \cite{Spielmann}.
 Bernardini (\cite{Bernardini}) investigated the transmission of wavepackets with energies above the height of the barrier.
He also finds that there is not only one reflected and one transmitted wave
packet, but many of them that leave the barrier. This result was obtained by 
the decomposition
of the transmission and reflection coefficients of the stationary solution in  
 infinite series.  Our result is the counterpart  in the tunneling regime. 

\section{The solution of the time-dependent Schr\"odinger equation}

A finite barrier 
 is described by the potential

\[
\begin{array}{lll}
V(x)=0&\qquad \mbox{for} \qquad & x\,\leq\,0\\
V(x)=V &\qquad \mbox{for} \qquad &0\,< x\,<\,d
\\  V(x)=0 &\qquad  \mbox{for} \qquad & x\,\geq\,d \quad.
 \end{array}
\]

The dynamics of a wave packet $ \psi(x,0) $  is determined by the Schr\"odinger equation 

\begin{align}
\label{Ba}
&-\frac{\hbar^2}{2 m} \frac{\partial^2 \psi(x,t)}{\partial x^2}
 = i  \hbar \,   \frac{\partial \psi(x,t)}{\partial t} \qquad &
 \mbox{for} \qquad x\,<\,0 \,,\\
& -\frac{\hbar^2}{2 m} \frac{\partial^2 \psi(x,t)}{\partial x^2}+V \psi(x,t)
 = i  \hbar \,   \frac{\partial \psi(x,t)}{\partial t} \qquad &
 \mbox{for} \qquad  0\,<\,x\,<\,d\,\,,\\
 &-\frac{\hbar^2}{2 m} \frac{\partial^2 \psi(x,t)}{\partial x^2}
 = i  \hbar \,   \frac{\partial \psi(x,t)}{\partial t} \qquad &
 \mbox{for} \qquad x\,>\,d \,,
 \end{align}
 
 where $\psi(x,t)$ is supposed to be continuously differentiable everywhere and square integrable .

We will further assume that the initial wave packets are located
at the left side of the barrier

 \begin{equation}
 \label{left}
\psi(x,0)=0 \qquad \mbox{for}\qquad x \geq 0\,.
 \end{equation}

We apply the method of Laplace transformation, already introduced in (\cite{Methode}). 

The Laplace transformed wavepacket $\varphi(x,s)$

\begin{equation}
\varphi(x,s) = \mathcal{L}(\psi(x,t)) = \int\limits_{0}^{\infty} \psi(x,t)  e^{-s t} dt \quad .
\end{equation}
\begin{subequations}
\label{Bat}
obeys the transformed equations
\begin{align}
& -\frac{\hbar^2}{2 m} \frac{\partial^2 \varphi(x,s)}{\partial x^2}  \,
\varphi(x,s) = i \hbar \, s \,  \varphi(x,s) - i \hbar \, \psi(x,0)\qquad &
 \mbox{for} \qquad x\,<\,0 \,,\\
&-\frac{\hbar^2}{2 m} \frac{\partial^2 \varphi(x,s)}{\partial x^2} + V \,
\varphi(x,s) = i \hbar \, s \,  \varphi(x,s) - i \hbar \, \psi(x,0) \qquad &
 \mbox{for} \qquad  0\,<\,x\,<\,d\,\,,\\
 &-\frac{\hbar^2}{2 m} \frac{\partial^2 \varphi(x,s)}{\partial x^2}  \,
\varphi(x,s) = i \hbar \, s \,  \varphi(x,s) - i \hbar \, \psi(x,0) \qquad &
 \mbox{for} \qquad x\,>\,d \,.
 \end{align}
 \end{subequations}
 The solution of (\ref{Bat}) is determined by the method of variation of constants:

\begin{subequations}
\label{LoesungBs}
\begin{align}
\varphi(x,s)&=\sqrt{\frac{m}{2 i s \hbar}} \left \{u_{1}(x,s)\int_{-d}^{x}u_{2}(y,s) 
\psi(y,0) dy \,  - \,\right.&
\left.
 u_{2}(x,s) 
\int_{-d}^{x}u_{1}(y,s) \psi(y,0) dy 
\right\} \nonumber \\
& &\nonumber \\
&\quad +\, \alpha(s) u_{1}(x,s)  \,+\, \beta(s) u_{2}(x,s) 
 &\mbox{for}  \quad x\,<\,0 \qquad\\
& &\nonumber \\
\varphi(x,s)&= \gamma(s) u_{3}(x,s)  \,+\, \delta(s) u_{4}(x,s)
 &\mbox{for} \quad 0\,<\, x\,<\,d \qquad \\
\varphi(x,s)&= \mu(s) u_{1}(x,s)  \,+\, \nu(s) u_{2}(x,s)
 &\mbox{for} \quad x\,>\,d. \qquad \,
\end{align}
\end{subequations}

The functions $u_{1}(x,s),u_{2}(x,s),u_{3}(x,s),u_{4}(x,s)$ are the solutions of the homogeneous
equations corresponding to (\ref{Bat})
\begin{subequations}
\label{u1234}
\begin{align}
& u_{1}(x,s)=e^{i\sqrt{\frac{2 m s i}{\hbar}} x}\quad 
& u_{2}(x,s)=e^{-i \sqrt{\frac{2 m s i}{\hbar}} x} &\mbox{for}\quad 0\,<\, x\,<\,d
\\
& u_{4}(x,s)=e^{-i \sqrt{\frac{2 m s i}{\hbar}-\frac{2 m V}{\hbar^2}} x}\,
& u_{3}(x,s)=e^{i\sqrt{\frac{2 m s i}{\hbar}-\frac{2 m V}{\hbar^2}} x}  &\mbox{for}\quad x\,<\, 0,x\,>\,d.
\end{align}
\end{subequations}

Since $\varphi(x,s)$ must vanish for $x\rightarrow \pm \infty$, we find
\[
 \alpha(s)=0\,,\quad \nu(s)=0\,.
\]

If we evaluate $\varphi(x,s)$ and its first derivative at $x=0$ and $x=d$, we obtain, imposing continuous differentiability

\begin{align}
&\beta(s)=\sqrt{\frac{m}{2 i \hbar s}}\left(I_{1}-I_{2}\right)
+\sqrt{\frac{m}{2 i \hbar s}}\cdot
\frac{2\, I_{2}}{1+\sqrt{1-\frac{V}{i \hbar s}}}\cdot
\frac{\rho(s)-e^{-2 i d \sqrt{\frac{2 m s i}{\hbar}-\frac{2 m V}{\hbar^2}}}}{\rho^2(s)-e^{-2 i d \sqrt{\frac{2 m s i}{\hbar}-\frac{2 m V}{\hbar^2}}}} \\
&\gamma(s)=
-\sqrt{\frac{m}{2 i \hbar s}}\cdot
\frac{2\, I_{2}}{1+\sqrt{1-\frac{V}{i \hbar s}}}\cdot
\frac{e^{-2 i d \sqrt{\frac{2 m s i}{\hbar}-\frac{2 m V}{\hbar^2}}}}{\rho^2(s)-e^{-2 i d \sqrt{\frac{2 m s i}{\hbar}-\frac{2 m V}{\hbar^2}}}}\\
&\delta(s)=\sqrt{\frac{m}{2 i \hbar s}}\cdot\frac{2\, I_{2}\,\rho(s)}{1+\sqrt{1-\frac{V}{i \hbar s}}}\cdot
\frac{1}{\rho^2(s)-e^{-2 i d \sqrt{\frac{2 m s i}{\hbar}-\frac{2 m V}{\hbar^2}}}} \,,\\
& \mu(s)=\sqrt{\frac{m}{2 i \hbar s}}\cdot
\frac{2\, I_{2}}{1+\sqrt{1-\frac{V}{i \hbar s}}}
e^{- i d \sqrt{\frac{2 m s i}{\hbar}}}\cdot
\frac{1-\rho(s)}
{\rho^2(s)-e^{-2 i d \sqrt{\frac{2 m s i}{\hbar}-\frac{2 m V}{\hbar^2}}}}
e^{-2 i d \sqrt{\frac{2 m s i}{\hbar}-\frac{2 m V}{\hbar^2}}}\,,
\end{align}
where we have introduced the abbreviations
\[
I_{1}=\int_{-\infty}^{0}e^{i \sqrt{\frac{2 m s i}{\hbar}}y}\psi(y,0)dy\,,
\quad
I_{2}=\int_{-\infty}^{0}e^{-i \sqrt{\frac{2 m s i}{\hbar}}y}\psi(y,0)dy
\]
and also used
\begin{equation}
 \rho(s)=\frac{2}{ 1+\sqrt{1-\frac{V}{\hbar s i}} }-1\,.
\end{equation}

Inserting this result into (\ref{LoesungBs}) and applying the  series expansion 
\[
 \frac{1}{1-\rho^2(s)e^{2 i d \sqrt{\frac{2 m s i}{\hbar}-\frac{2 m V}{\hbar^2}}}}=\sum_{k=0}^{\infty}
 (\rho(s))^{2 k}e^{2 i d k\sqrt{\frac{2 m s i}{\hbar}-\frac{2 m V}{\hbar^2}}}
 \]
yields 
\begin{subequations}
\label{Tunnelabc}
\begin{align}
& \varphi(x,s)=
 \sqrt{\frac{m}{2 s i \hbar}}\left\{
\int_{-\infty}^{0}
e^{i\sqrt{\frac{2 m s i}{\hbar}} |x-y|} \psi(y,0) dy+ 
\int_{-\infty}^{0}
e^{-i\sqrt{\frac{2 m s i}{\hbar}} (x+y)}\psi(y,0) dy
\cdot \rho(s)\right\} + \nonumber     \\
& \qquad \sqrt{\frac{m}{2 s i \hbar}}
\sum_{l=0}^{\infty}
\int_{-\infty}^{0}
e^{2 i d (l+1)\sqrt{\frac{2 m s i}{\hbar}-\frac{2 m V}{\hbar^2}}
-i (x+y)\sqrt{\frac{2 m si}{\hbar}}}\psi(y,0)dy
\cdot (\rho^2(s)-1)\rho^{2l+1}(s)
\nonumber \\
& \qquad \mbox{for}  \quad x\,<\,0 \label{Tunnela}\\
& \varphi(x,s)=
 \sqrt{\frac{m}{2 s i \hbar}}
  \sum_{l=0}^{\infty}
\int_{-\infty}^{0}
e^{ i  (2 d l+x)\sqrt{\frac{2 m s i}{\hbar}-\frac{2 m V}{\hbar^2}}
-i y \sqrt{\frac{2 m si}{\hbar}}}\psi(y,0)dy
\cdot (\rho(s)+1)\rho^{2l}(s)\,-\nonumber\\
& \qquad \sqrt{\frac{m}{2 s i \hbar}}
  \sum_{l=0}^{\infty}
\int_{-\infty}^{0}
e^{ i  (2 d (l+1)-x)\sqrt{\frac{2 m s i}{\hbar}-\frac{2 m V}{\hbar^2}}
-i y\sqrt{\frac{2 m si}{\hbar}}}\psi(y,0)dy
\cdot (\rho(s)+1)\rho^{2l+1}(s)\,\nonumber\\
& \qquad \mbox{for}  \quad 0\,\leq x\,\leq \,d \,,\label{Tunnelb}\\
& \varphi(x,s)=
 \sqrt{\frac{m}{2 s i \hbar}}
\sum_{l=0}^{\infty}
\int_{-\infty}^{0}
e^{i d (2l+1)\sqrt{\frac{2 m s i}{\hbar}-\frac{2 m V}{\hbar^2}}
+i (x-d-y)\sqrt{\frac{2 m si}{\hbar}}}\psi(y,0)dy
\cdot (-\rho^2(s)+1)\rho^{2l}(s) \nonumber \\
& \qquad \mbox{for}  \quad x\,>\,d \label{Tunnelc}\,.
\end{align}
\end{subequations}

We introduce the shifted momentum representation 
\begin{equation}
 f(Q,p)=\frac{1}{\sqrt{2 \pi \hbar}}\int_{-\infty}^{\infty} \psi(x+Q,0) e^{\frac{-i p x}{\hbar}} dx\,=\,
 e^{\frac{i p Q}{\hbar}}f(p)\,,
\end{equation}
where 
\[
 f(0,p) = f(p)
\]
denotes the representation of the wave function in momentum space.

Applying the abbreviations

\begin{align*}
& a_{l}(t)=\mathcal{L}^{-1}\left\{\rho^{2 l+1}(s)(1-\rho^2(s))\right\}\,, \quad
b_{l}(t)=\mathcal{L}^{-1}\left\{\rho^{2 l}(s)(1+\rho(s))\right\} \\
& c_{l}(t)=\mathcal{L}^{-1}\left\{\rho^{2 l+1}(s)(1+\rho(s))\right\}\,, \quad
g_{l}(t)=\mathcal{L}^{-1}\left\{\rho^{2 l}(s)(1-\rho^2(s))\right\}\,,
\end{align*}
we find proceeding as for the asymmetric square well in \cite{Methode} 
 for the inverse Laplace transform of (\ref{Tunnelabc})
\begin{subequations}
\label{TunnelL}
\begin{align}
&\psi(x,t)=
\frac{\kappa}{\sqrt{2 \pi t \,i}} 
\int_{-\infty}^{0}  e^{\frac{i (x-y)^2 \kappa ^2}{2 t}}\psi(y,0) dy
+
\int_{0}^{t}\frac{\kappa}{\sqrt{2 \pi (t-\tau) \,i}} 
\int_{-\infty}^{0}e^{\frac{i (x+y)^2 \kappa ^2}{2 (t-\tau)}} \psi(y,0) r(\tau)dy\, d\tau\,+ \nonumber \\
& \quad\sum_{l=0}^{\infty}\int_{0}^{t}\int_{-\infty}^{\infty}
K\left(2d( l+1),p,t-\tau\right)
f(-x,p)dp \cdot a_{l}(\tau)d\tau \nonumber\\
& \quad \mbox{for}  \quad x\,<\,0 \,, \label{TunnelLa}\\
&\psi(x,t)=
\sum_{l=0}^{\infty}\int_{0}^{t}\int_{-\infty}^{\infty}
K\left( 2 d l+x),p,t-\tau\right)
f(p)dp \cdot b_{l}(\tau)d\tau\,+ \nonumber \\
& \quad\sum_{l=0}^{\infty}\int_{0}^{t}\int_{-\infty}^{\infty}
K\left(d (2l+1)-x),p,t-\tau\right)
f(p)dp \cdot c_{l}(\tau)d\tau\, \nonumber\\
& \quad \mbox{for}  \quad 0\,<\,x\,<\,d \,, \label{TunnelLb}\\
&\psi(x,t)=
\sum_{l=0}^{\infty}\int_{0}^{t}\int_{-\infty}^{\infty}
K\left(d(2 l+1),p,t-\tau\right)
f(x-d,p)dp \cdot g_{l}(\tau)d\tau\,\nonumber\\
& \quad \mbox{for}  \quad x\,>\,d \,, \label{TunnelLc}
\end{align}
\end{subequations}
where $K(x,p,t)$ is defined by
\begin{align}
& K(x,p,t)= \frac{1}{2 \sqrt{2 \pi \hbar}} e^{\frac{- i V t}{\hbar}-\frac{i t q^2}{2 \hbar m}}
 \quad \cdotp
 \nonumber \\
 & \quad 
 \left\{
 e^{-\frac{i x q}{\hbar}} \mbox{Erfc}
 \left[-i \sqrt{\frac{2 m i}{\hbar\, t}} \frac{x}{2}-i \sqrt{\frac{i\, t}{2 \hbar m}} q\right]
 +e^{\frac{i x q}{\hbar}} \mbox{Erfc}
 \left[-i \sqrt{\frac{2 m i}{\hbar\, t}} \frac{x}{2}+i \sqrt{\frac{i\, t}{2 \hbar m}} q\right]
\right\} \,,
 \nonumber \\
 \end{align} 
 and we used $\kappa= \sqrt{\frac{m}{\hbar}}$ and $q=\sqrt{p^2-2 m V}$.
 
\section{Tunneling of wave packets}

In order to investigate the tunneling process, we
 consider  a wave packet that is represented in momentum space by 
 \begin{subequations}
\label{Darstellung}
\begin{equation}
 f(p)=e^{\frac{-i p x_{0}}{\hbar}} F\left(p-p_{0}\right) \quad,
\end{equation}

where $F(p)$ fulfills
\begin{equation}
 \int_{-\infty}^{\infty} F^{*}(p)\,p\, F(p) dp=0 \,,\qquad
  \int_{-\infty}^{\infty} \, F^{*}(p)F'(p) dp=0\,.
\end{equation}
\end{subequations}.
The expectation values of position and momentum are then given by
\[
\left\langle\hat{x}\right\rangle=x_{0}\,,
\qquad
\left\langle\hat{p}\right\rangle=p_{0}\,.
\]
We  assume that the wavepacket is  concentrated within a region $0\,<\,p\,<p_{max}\,<\sqrt{2 m V}$.
so that we can use the approximation
\begin{equation}
\label{Rand2}
 f(p) \approx 0 \quad 
 \mbox{for} 
 \quad p\,>\,p_{max} \quad \mbox{and}
\quad  p\,<\,p_{min}\,.
\end{equation}
Moreover it should be sufficiently peaked around the momentum expectation value to justify the approximation
\begin{equation}
\label{Approxp1}
F(p-p_{0}) (p-p_{0})
 \approx 0\,.
 \end{equation}
Finally the difference $\sqrt{2 m V}-p_{m}$ should be big enough
to ensure 
\begin{equation}
\label{Faktor2}
 \frac{1}{\sqrt{1-\frac{p^2}{2 m V}}} = O(1)\quad \mbox{for} 
 \quad p_{min} \,\leq\, p\,\leq\,p_{max}\,.
\end{equation}
In Appendix~\ref{Consistency} it is shown that these assumptions about the momentum distribution are compatible with the requirement (\ref{left})
for $\psi(x,0)$.

We start with the solution for $x>d$ (\ref{Tunnelc}).
If we rewrite $K(x,p,t)$  for $q^2=p^2-2 m V\,<\,0$ and use 
\[
 X_{l}\equiv d(2l+1)\,,
\]
 we find  (see section 4.4 in \cite{Methode} for details)

\begin{align}
&K(X_{l},p,t)=U_{0}+U_{1}-U_{2} \,,\quad \mbox{where} \quad \\
  & U_{0}=e^{-\frac{i V t}{\hbar}}
\frac{1}{\sqrt{2 \pi \hbar}}
 e^{-\frac{i t q^2}{2 \hbar m}+\frac{i X_{l} q}{\hbar}} =
 \frac{1}{\sqrt{2 \pi \hbar}}
 e^{- \frac{i p^2 t}{2 m \hbar}-\frac{X_{l} \sqrt{2 m V-p^2}}{\hbar}}\,,
\nonumber \\
& U_{1}=
 \frac{\kappa}{ \sqrt{2 \pi t \,i}} \frac{1}{\sqrt{2 \pi \hbar }}
e^{-\frac{i V t}{\hbar}}\int_{0}^{\infty}  e^{\frac{i (X_{l}+u)^2 \kappa ^2}{2 t}}
e^{\frac{i \sqrt{2 m V-p^2} u}{\hbar}}
du\,, \\
&U_{2}=\frac{\kappa}{ \sqrt{2 \pi t \,i}} \frac{1}{\sqrt{2 \pi \hbar }}
e^{-\frac{i V t}{\hbar}}\int_{0}^{\infty}  e^{\frac{i (X_{l}-u)^2 \kappa ^2}{2 t}}
e^{\frac{i \sqrt{2 m V-p^2} u}{\hbar}}du
\, . \nonumber
 \end{align}
Inserting 
\[
  \frac{\kappa}{ \sqrt{2 \pi t \,i}}e^{\frac{i (X_{l}\pm u)^2 \kappa ^2}{2 t}}=
  \int _{-\infty}^{\infty}
  \frac{1}{2 \pi \hbar}
  e^{\frac{-i q^2 t}{2 m \hbar}+\frac{i q X_{l}}{\hbar}\pm \frac{i q u}{\hbar}}dq\,,
\]
we obtain
\begin{equation}
 U_{1}-U_{2}=
 \frac{1}{(2 \pi \hbar)^{3/2}}e^{-\frac{i V t}{\hbar}}
 \int_{-\infty}^{\infty} 
 e^{\frac{-i q^2 t}{2 m \hbar}+\frac{i q X_{l}}{\hbar}}
 \frac{2 i \hbar q}{2 m V-p^2+q^2}dq \,.
\end{equation}

 We find for  $U_{1}-U_{2}$, that contains no oscillatory part
 \begin{equation}
 \label{Schnitt}
 \int_{-\infty}^{\infty}(U_{1}-U_{2})f(x-d,p)dp\approx 0\,,
  \end{equation}
 since
 \begin{align*}
  \int_{-\infty}^{\infty}f(p) p^{n} e^{i \frac{p (x-d)}{\hbar}} dp= 
  \left(-i \hbar \frac{\partial}{\partial x }\right)^n\psi(x-d,0)=0\quad \mbox{for} \quad\, x\,>\,d\,,\,n=0,1,2,N\,,
 \end{align*}
 and so the contribution to (\ref{Schnitt}) will be proportional to
 $\Delta p^{N+1}$\,,
 if $f(p)$ decays rapidly enough so that the integrals
 \[
 \int_{-\infty}^{\infty}f(p) p^n dp
  \]
  exist up to $n=N+1$.

 If $\psi(x)$ is infinitely differentiable for
 $x>d$, th eleft hand side of (\ref{Schnitt}) vanishes identically.
So we conclude that for wavepackets that are sufficiently smooth for $x\,>\,d$ , $U_{0}$ is the only relevant contribution.

We obtain for the wavefunction for $x\,>\,d$

\begin{align}
 &\psi(x,t) \approx
 \sum_{l=0}^{\infty}\int_{0}^{t}\frac{1}{\sqrt{2 \pi \hbar}}
 e^{-\frac{X_{l} \sqrt{2 m V-p_{0}^2}}{\hbar}}\int_{-\infty}^{\infty}e^{- \frac{i p^2 (t-\tau)}{2 m \hbar}}
 e^{\frac{i p (x-d)}{\hbar}}
 f(p) dp \cdot g_{l}(\tau)d\tau \label{Resultata}\,.
\end{align}

This result contains the free time evolution  of the initial wave packet
that is shifted by $d$ to the right. Proceeding as for the asymmetric square well(see \cite{Methode}) we can approximate the 
convolution integral by
\begin{align}
&\int_{0}^{t}\frac{1}{\sqrt{2 \pi \hbar}}
 \int_{-\infty}^{\infty}e^{- \frac{i p^2 (t-\tau)}{2 m \hbar}}
 e^{\frac{i p (x-d)}{\hbar}}
 f(p) dp \cdot g_{l}(\tau)d\tau \approx \label{ConvI}\\
 &\int_{0}^{\infty}\frac{1}{\sqrt{2 \pi \hbar}}
 \int_{-\infty}^{\infty}e^{- \frac{i p^2 (t-\tau)}{2 m \hbar}}
 e^{\frac{i p (x-d)}{\hbar}}
 f(p) dp \cdot g_{l}(\tau)d\tau \approx \nonumber \\
 &\frac{1}{\sqrt{2 \pi \hbar}}
 \int_{-\infty}^{\infty}e^{- \frac{i p^2 t}{2 m \hbar}}
 e^{\frac{i p (x-d)}{\hbar}} R(p)^{2l} (1-R(p)^2)f(p) dp 
 \, \label{TIntegralH}\\
 & \approx R(p_{0})^{2l} (1-R(p_{0})^2) \cdot \frac{1}{\sqrt{2 \pi \hbar}}
 \int_{-\infty}^{\infty}e^{- \frac{i p^2 t}{2 m \hbar}}
 e^{\frac{i p (x-d)}{\hbar}}f(p) dp \,, \label{TIntegral}
\end{align}
where $R(p)$ is given by 
\begin{equation}
 \label{Rk}
 R(p)=-1 + 2 k-2 \sqrt{k(k-1)}
 \quad \mbox{with} \quad k=\frac{p^2}{2 m V}\,.
\end{equation}
Here the assumptions about the concentration
of the wave packet (\ref{Approxp1}, \ref{Faktor2}) justifies putting $R(p)$ before 
the integral. Evaluating the sum in (\ref{Resultata}), we obtain
\begin{align}
\label{PsiTunnel}
\psi(x,t) \approx \sum_{l=0}^{\infty}\int_{0}^{t}\frac{1}{\sqrt{2 \pi \hbar}}
 e^{-\frac{X_{l} \sqrt{2 m V-p_{0}^2}}{\hbar}}\int_{-\infty}^{\infty}e^{- \frac{i p^2 (t-\tau)}{2 m \hbar}}
 e^{\frac{i p (x-d)}{\hbar}}
 f(p) dp \cdot g_{l}(\tau)d\tau \approx\nonumber \\
 \frac{e^{\frac{- d \sqrt{2 m V-p_{0}^2}}{\hbar}}}{1-e^{\frac{-2 d \sqrt{2 m V-p_{0}^2}}{\hbar}}R^2(p_{0})}
 (1-R^2(p_{0}))
 \frac{1}{\sqrt{2 \pi \hbar}}
 \int_{-\infty}^{\infty}e^{- \frac{i p^2 t}{2 m \hbar}}
 e^{\frac{i p (x-d)}{\hbar}}f(p) dp \,.
 \end{align}
For the neglected part of the time integral (\ref{ConvI})
\begin{equation}
\label{Delta}
\Delta_{l}\equiv \int_{t}^{\infty}\frac{1}{\sqrt{2 \pi \hbar}}
 \int_{-\infty}^{\infty}e^{- \frac{i p^2 (t-\tau)}{2 m \hbar}}
 e^{\frac{i p (x-d)}{\hbar}}
 f(p) dp \cdot g_{l}(\tau)d\tau \,,
 \end{equation}
we find the estimates (see Appendix \ref{IE})
\begin{equation}
\left|\Delta_{l}\right|\sim O(l)\cdot 
O\left(\left[\frac{2 \hbar}{V t}\right]^{\frac{1}{4}}\right)
\end{equation}
Since 
\[
\sum_{l=0}^{\infty} 
e^{-\frac{X_{l} \sqrt{2 m V-p_{0}^2}}{\hbar}}\Delta_{l}
\sim \frac{e^{\frac {d \sqrt{2 m V-p_{0}^2}}{\hbar}}}
{\left(-1+e^{\frac{2 d \sqrt{2 m V-p_{0}^2}}{\hbar}}\right)^2}\cdot 
O\left(\left[\frac{2 \hbar}{V t}\right]^{\frac{1}{4}}\right)
\]
the approximation for $\psi(x,t)$ given by (\ref{PsiTunnel})
can be used for $t \gg \frac{\hbar}{V}$. 

The wavepacket \ref{PsiTunnel} describes a free evolving wavepacket that started at $t=0$ at the position $x_{0}+ d$. It leaves the 
barrier without any time delay and is instantaneously transmitted through the barrier attenuated 
by a factor of the magnitude
\begin{equation}
  e^{\frac{ -d \sqrt{2 m V-p_{0}^2}}{\hbar}}\,.
\end{equation}

So within our approximations we find that the tunneling time is zero, where we assume in accordance with \cite{Hartmann} 
that the transmission begins when a freely evolving 
wave packet starting at $t=0$ at the position $x_{0}$ would have arrived at the barrier, or equivalently when a classical
particle starting at $t=0, x_{0}=0$ has reached the barrier. It is not practicable to follow the peak of the original wavepacket until the 
entrance of the tunnel since this peak will be deformed by oscillations during the reflection process(\cite{Methode},\cite{LosTunnel}) and also the position expectation value
might undergo a slight attenuation before the tunnel as it was shown for infinite walls (\cite{Bounce}).

Performing the inverse Laplace transform for $x<0$ and $0\,<\,x\,<\,d$ 
(\ref{Tunnela},\ref{Tunnela})and making the same approximations 
as in the previous case we find for the wavepacket
to the left of and within the barrier:

\begin{align}
 & \psi (x,t)\approx \frac{\kappa}{\sqrt{2 \pi t \,i}} 
\int_{-\infty}^{0}  e^{\frac{i (x-y)^2 \kappa ^2}{2 t}}\psi(y,0) dy
+
  \frac{1}{\sqrt{2 \pi \hbar}}
 \int_{-\infty}^{\infty}e^{- \frac{i p^2 t}{2 m \hbar}}
 e^{\frac{-i p x}{\hbar}} R(p)f(p) dp + \nonumber\\
 & \sum_{l=0}^{\infty}
 e^{-\frac{2 d (l+1) \sqrt{2 m V-p_{0}^2}}{\hbar}}\frac{1}{\sqrt{2 \pi \hbar}}
 \int_{-\infty}^{\infty}e^{- \frac{i p^2 t}{2 m \hbar}}
 e^{\frac{-i p x}{\hbar}} R(p)^{2 l +1}(R(p)^2-1)f(p) dp \approx \nonumber \\
&  \frac{\kappa}{\sqrt{2 \pi t \,i}} 
\int_{-\infty}^{0}  e^{\frac{i (x-y)^2 \kappa ^2}{2 t}}\psi(y,0) dy +  \nonumber\\
& \left\{1+\frac{e^{\frac{- 2 d \sqrt{2 m V-p_{0}^2}}{\hbar}}}
{1-e^{\frac{-2 d \sqrt{2 m V-p_{0}^2}}{\hbar}}R(p_{0})^2} \cdot
(R^2(p_{0})-1)
 \right\}
  R(p_{0})\frac{1}{\sqrt{2 \pi \hbar}}
 \int_{-\infty}^{\infty}e^{- \frac{i p^2 t}{2 m \hbar}}
 e^{\frac{-i p x}{\hbar}} f(p) dp \nonumber\\
 & \mbox{for} \,x\,<\,0\, ,\label{PsiReflex}
\end{align}
\begin{align}
 & \psi (x,t)\approx  \sum_{l=0}^{\infty}
 e^{-\frac{(2 d l+x) \sqrt{2 m V-p_{0}^2}}{\hbar}}
\frac{1}{\sqrt{2 \pi \hbar}} \int_{-\infty}^{\infty}e^{- \frac{i p^2 t}{2 m \hbar}}
  R(p)^{2 l}(R(p)+1)f(p) dp \approx \nonumber \\
 &\,-\sum_{l=0}^{\infty}
 e^{-\frac{(2 d (l+1)+x) \sqrt{2 m V-p_{0}^2}}{\hbar}}
\frac{1}{\sqrt{2 \pi \hbar}} \int_{-\infty}^{\infty}e^{- \frac{i p^2 t}{2 m \hbar}}
 R(p)^{2 l+1}(R(p)+1)f(p) dp \approx \nonumber \\
& \left\{\frac{1}
{1-e^{\frac{-2 d \sqrt{2 m V-p_{0}^2}}{\hbar}}R(p_{0})^2} \,
-\,
 \frac{e^{\frac{- 2( d-x) \sqrt{2 m V-p_{0}^2}}{\hbar}}}
{1-e^{\frac{-2 d \sqrt{2 m V-p_{0}^2}}{\hbar}}R(p_{0})^2} \cdot
 R(p_{0})
 \right\} \nonumber \\
&\,\cdot (R(p_{0})+1) e^{\frac{- x\sqrt{2 m V-p_{0}^2}}{\hbar}}\cdot\,
\frac{1}{\sqrt{2 \pi \hbar}}
 \int_{-\infty}^{\infty}e^{- \frac{i p^2 t}{2 m \hbar}}
  f(p) dp \nonumber\\
 & \mbox{for} \,0\,<\,x\,<\,d\, .\label{PsiBarrire}
\end{align}

The solution within the barrier differs from the solution of the potential step (see \cite{Methode}) only
by a time-independent factor

\begin{equation}
\label{Psistep}
\psi_{step}(x,t)\approx (1+R(p_{0}))
e^{-\frac{ \sqrt{2 m V-p_{0}^2} x}{\hbar}}
\frac{1}{\sqrt{2 \pi \hbar}} 
\int_{-\infty}^{\infty}  e^{\frac{- i p^2 t}{2 m \hbar}}
f(p) dp\,.
\end{equation}

So we see that the time it
needs until the barrier is (approximately) empty again, is
independent of the thickness of the barrier.
The solution on the left side consists of the incoming and the reflected wave packet (\ref{PsiReflex}).
In contrast to the potential step the reflected wavepacket experiences a permanent
attenuation, since a part of the wavefunction has tunneled through the barrier.
After the reflection process the wavefunction consists of a reflected and a transmitted
wavepacket (\ref{PsiTunnel}) only.
An explicit calculation yields

\begin{align}
&\left| \left\{1+\frac{e^{\frac{- 2 d \sqrt{2 m V-p_{0}^2}}{\hbar}}}
{1-e^{\frac{-2 d \sqrt{2 m V-p_{0}^2}}{\hbar}}R(p_{0})^2} \cdot
(R^2(p_{0})-1)
 \right\}
  R(p_{0}) \right|^2+ \\
&  \left|\frac{e^{\frac{- d \sqrt{2 m V-p_{0}^2}}{\hbar}}}{1-e^{\frac{-2 d \sqrt{2 m V-p_{0}^2}}{\hbar}}R^2(p_{0})}
 (1-R^2(p_{0}))\right|^2\,=\,1 \,,
\end{align}

which confirms that the integral over the probability density is conserved and our approximations
are consistent.
\section{The tunneling time}

Within our approximations we found the tunneling time to be zero, 
since our solution (\ref{PsiTunnel})
indicates that the wavepacket leaves the tunnel, shifted by d to the right. 
Here we have assumed, 
that all functions of $R(p)$ can be pulled out of the integral (\ref{TIntegralH}), and 
therefore 
they do not influence the 
dynamics of the wave packet. If we take into account the first order contributions of $R(p)$, 
we find
a small, but finite tunneling time. 
We will from now on assume that the initial wave function
is an uncorrelated function of the form
\[
 f(p)=G(p)e^{\frac{-i p x_{0}}{\hbar}} \,,
\]
where $G(p)$ is a real function that yields a momentum expectation value $p_{0}$. The position expectation value is then given by $x_{0}$.

The impact of an additional factor $Z(p)$ on the initial wavepacket $f(p)$ is twofold
\begin{equation}
\label{Z}
 Z(p)f(p)=z(p)e^{i \mu (p)} f(p)\,,\mbox{where}\quad z(p)=|Z(p)|\,.
\end{equation}

The probability density in momentum space is only affected by the absolute value $z(p)$.
If we restrict ourselves to first order contributions in $p-p_{0}$, 
\[
 z(p) \approx z(p_{0})+ z_{1}(p-p_{0})
\]

we find that we can neglect
the momentum shift:
\begin{align}
 &N^2\equiv\int_{-\infty}^{\infty}|f(p)|^2 z(p)^2 dp
 \approx  z(p_{0})^2\int_{-\infty}^{\infty}|f(p)|^2 dp\,=
 z(p_{0})^2 \\
& \left\langle \hat{p}\right\rangle\,=\,
N^{-2}\int_{-\infty}^{\infty}p|f(p)|^2 z(p)^2 dp \\
&\approx
p_{0}+2 (z(p_{0 }))^{-1} z_{1}\int_{-\infty}^{\infty}|f(p)|^2 p(p-p_{0}) dp \approx
p_{0}\,.
 \end{align}
 Using also a linear approximation  for $\mu'(p)$
 \[
  \mu'(p) \approx \mu_{1}+\mu_{2}(p-p_{0})\,
 \]
we find for the position expectation value
 \begin{align}
  &\widetilde{x_{0}} =
  N^{-2}\int_{-\infty}^{\infty} z(p) f^{*}(p)e^{-i \mu (p)}
  \left(i \hbar \frac{\partial}{\partial p}\right)
  z(p) f(p)e^{i \mu (p)} dp = \\
 &N^{-2} x_{0}\int_{-\infty}^{\infty}G(p)^2 z(p)^2 dp
 -N^{-2} \hbar\int_{-\infty}^{\infty}G(p)^2 z(p)^2 \mu'(p)dp\\
 \approx
 & x_{0}-\hbar \mu_{1}\,.
  \end{align}

Therefore the phase of the functions
\begin{equation}
  R(p)^{2l} (1-R(p)^2)
\end{equation}
in (\ref{TIntegralH})
yields a shift of the position of each particular wave packet constituting  (\ref{PsiTunnel})
\begin{equation}
\label{PsiTunnel2}
 \psi(x,t) \approx
 \sum_{l=0}^{\infty}\frac{1}{\sqrt{2 \pi \hbar}}e^{-\frac{X_{l} \sqrt{2 m V-p_{0}^2}}{\hbar}}
 \int_{-\infty}^{\infty}e^{- \frac{i p^2 t}{2 m \hbar}}
 e^{\frac{i p (x-d)}{\hbar}} R(p)^{2l} (1-R(p)^2)f(p) dp \,.
\end{equation}

We find
\begin{align}
&Arg\left[1-R(p)^2\right]=
Arctan\left[\frac{-1+2 l}{2 \sqrt{k-k^2}}\right] \approx 
Arctan\left[\frac{-1+2 k_{0}}{2 \sqrt{k_{0}-k_{0}^2}}\right]+
\frac{2}{\sqrt{2 m V-p^2_{0}}}\cdot(p-p_{0}) \label{Phase1}\\
&Arg\left[R(p)\right]=
-Arccos\left[1-2 k\right] \approx 
-Arccos\left[1-2 k_{0}\right]+
\frac{2}{\sqrt{2 m V-p^2_{0}}}\cdot(p-p_{0}) \label{Phase2}\,,
\end{align}
where we have used the definitions
\[
 k=\frac{p^2}{2 m V}\,,\quad k_{0}=\frac{p_{0}^2}{2 m V}\,.
\]
So we see that the lth term of (\ref{PsiTunnel}) will experience a phase shift of 
\[
 \frac{2 (1+2l) (p-p_{0})}{\sqrt{2 m V-p^2_{0}}}\,,
\]
corresponding to a translation by
\begin{equation}
\label{Shift}
\delta x_{l} = \frac{2 (1+2l) \hbar}{\sqrt{2 m V-p^2_{0}}}=\frac{2 (1+2l) \hbar}{p_{0}} \sqrt{\frac{k_{0}}{1-k_{0}}}
\end{equation}

to the left. So the delay time will be
\[
T_{l}=  \frac{2 (1+2l) m \hbar}{p_{0}\sqrt{2 m V-p^2_{0}}}
\]
instead of zero.

Each term is attenuated by a factor of the magnitude
\begin{equation}
\label{Abfall}
  e^{\frac{ -X_{l} \sqrt{2 m V-p_{0}^2}}{\hbar}}\,.
\end{equation}
For thick barriers, if
\[
 d \frac{\sqrt{2 m V-p_{0}^2}}{\hbar} \gg 1\,,
\]

the first term will dominate the sum (\ref{PsiTunnel}), and the tunneling time will be given by
\begin{equation}
\label{TTunnel}
 T_{0}=\frac{2 m \hbar}{\sqrt{2 m V-p^2_{0}}p_{0}}
\end{equation}

\begin{figure}[htp]
\includegraphics[width=0.7 \textwidth]{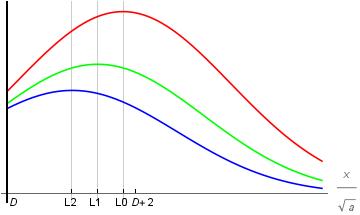} 
\caption{The  relative dimensionless probability density $|\psi|^2 \cdot (\pi a)^{1/2}$ of the first three wavepackets leaving 
a barrier with the width $d=D \cdot \sqrt{a}$. The initial wavepacket is of the form (\ref{initial}) with the choice of parameters 
$x_{0}=-20\cdot\sqrt{a}$, $p_{0}=10\cdot\hbar/\sqrt{a}$, $L=20,\,k_{0}=1/2$, where $L$ characterizes the detailed decreasing behaviour.
The uncertainties are determined
by (\ref{Varianz}):$(\Delta x) ^{2}=0.49\cdot a$, $(\Delta p) ^{2}=0.51 \cdot \hbar^2 \cdot a^{-1}$. 
The first three terms
of (\ref{PsiTunnel2}) were evaluated at the time $t=t_{R}+2\cdot \sqrt{a}\cdot m\cdot p_{0}^{-1}$ where 
$t_{R}=-x_{0}\cdot m \cdot p_{0}^{-1}$ denotes 
the time when the centre
of a freely evolving wavepacket starting at $x_{0}$ has reached the barrier. An instantaneously transmitted wavepacket
would then immediately be at position $D \cdot\sqrt{a}$ and therefore reach $\sqrt{a}\cdot( D+2)$ after a time interval 
$2 \sqrt{a}\cdot m\cdot p_{0}^{-1}$. 
The evaluation shows a slight delay according to (\ref{Shift}), so that the wavepackets are centred at 
$L0\cdot \sqrt{a} = D \cdot \sqrt{a}-2 \cdot \hbar \cdot p_0^{-1}$, 
$L1 \cdot \sqrt{a} = D \cdot \sqrt{a}-6 \cdot \hbar \cdot p_0^{-1}$, 
$L2 \cdot \sqrt{a} = D\cdot \sqrt{a}-10\cdot \hbar \cdot p_0^{-1}$.
Note that the width $d$ only matters with regard to the magnitude of the wavepackets and
apart from this the picture applies to a range of possible widths restricted only by the consistency condition (\ref{Bcondition}). 
Here we have chosen the magnitude of the wavepackets arbitrarily to 
provide a better visibility. The integrals were numerically evaluated with Wolfram Mathematica, so that the picture is an additional confirmation
of (\ref{Shift}). Due to the high concordance of the  wavefunction with its  extension along the real line we could use  this extended function 
for the calculations (see Appendix~\ref{Consistency}).
} 
  \label{Blubber}
\end{figure}

This is exact the time Hartmann \cite{Hartmann}found as upper limit for the tunneling 
time through
thick barriers.
If the barrier gets thinner the other wavepackets for $l>0$
will also come into play (see figure \ref{Blubber}). Each of them leaves the barrier at a different time $T_{l}$.
But since they appear very shortly after each other they may appear as one smeared out wavepacket
with a delay time bigger than $T_{0}$.

 Note that apart from the absolute and relative magnitude of the wavepackets, all characteristic
quantities of tunneling are independent of the width $d$. In the case of the wavefunction pictured in figure \ref{Blubber} where the uncertainties are
given by 
\begin{equation}
\Delta x\approx\sqrt{\frac{a}{2}}\,,\quad\Delta p \approx\sqrt{\frac{\hbar}{2 a}}\,,
\end{equation}
the ratio between the distance of the centre of the wavepackets and the position uncertainty the determines the distinguishability between the wavepackets reads
\begin{equation}
 \frac{\delta x_{l+1}-\delta x_{l}}{\Delta x}= 
 \frac{4 \hbar}{p_{0} \Delta x}\sqrt{\frac{k_{0}}{1-k_{0}}} \approx \frac{8 \Delta p}{p_{0}}\sqrt{\frac{k_{0}}{1-k_{0}}}\,.
\end{equation}

\section{Discussion and conclusions }
Our result for the  tunneling time (\ref{TTunnel}) is not an exact reproduction
of Hartmann's result (\cite{Hartmann},\cite{Winful}) which predicts an increasing tunneling time with the thickness of the barrier 
before saturation 
takes place. However Spielmann et al. found a decreasing tunneling time in their experiments with
electromagnetic waves propagating through photonic band gap materials (see \cite{Spielmann} fig. 3, \cite{Pereyra} fig 1 ).
This qualitative behaviour is in good agreement with our results that also predict a decreasing tunneling time since for
thicker barriers the later wave packets are more and more attenuated.


For thicker barriers the conclusion of both calculations is that the tunneling time for sufficiently
peaked wave packets is given by (\ref{TTunnel}).
This is as far interesting as the results were obtained by completely different methods. Moreover we ensured in our calculations
that the initial wavepacket is only located at the left side of the barrier (\ref{left}) which is not clearly
guaranteed by Hartmann's approach. So our result makes sure that the Hartmann time (\ref{TTunnel}) is not some relic of the parts of 
the initial wavepacket that were at the right hand side
of the barrier from the beginning. 

We did not take into account the parts of the initial wavepacket with energies 
near or greater than the critical energy of standard transmission through the barrier
(see \ref{Rand2}). In Appendix~\ref{Consistency} we have derived a consistency condition for this 
approximation for the case of a special class of initial wavepackets. So this is a further counterexample to the idea that 
only energies greater than the barrier height contribute to tunneling (\cite{Chiao}).

We also found out that the approximate solution within a finite barrier differs
from the solution within the potential step only by a time-independent factor (\ref{PsiBarrire},\ref{Psistep})
which also indicates that important
dynamical properties are independent of the thickness of the barrier.
It would be especially interesting if
this is also true for more general tunneling processes as the tunneling out of a potential well that could model
radioactive decay or tunneling out of atoms as provided by the attoclock experiment  \cite{Landsmann}. Moreover an application of
the method of Laplace transformation to relativistic wave equations would yield a picture of the reflection and tunneling processes
in the relativistic case.
\section*{Acknowledgments}

We thank the referees for their comments and suggestions that helped to improve this article.
\appendix
\section{Estimation of the convolution integral} 
\label{IE}

According to \cite{Erdely}, the inverse Laplace transform of $\rho(s)^l$ is given by 
\begin{equation}
\mathcal{L}^{-1} \left( \rho(s)^l\right)= 
\frac{l}{i^l t}J_{l}\left[\frac{V t}{2 \hbar}\right]e^{-\frac{i V t}{2 \hbar}}\,.
\end{equation}

Therefore we find
\begin{equation}
g_{l}(t)=\mathcal{L}^{-1} \left(\rho^(2 l)(s)(1-\rho^2(s))\right)=
\frac{l}{i^(2l) t}J_{2l}\left[\frac{V t}{2 \hbar}\right]e^{-\frac{i V t}{2 \hbar}}-
\frac{l}{i^(2(l+1)) t}J_{2(l+1)}\left[\frac{V t}{2 \hbar}\right]e^{-\frac{i V t}{2 \hbar}}\,.
\end{equation}

For an integral of the form
\begin{align}
\label{u}
u(l,t)=
\int_{t}^{\infty} e^{\frac{\mp i p Q}{\hbar}}
e^{-\frac{i p^2 (t-\tau)}{2m \hbar}} e^{-\frac{i V \tau}{2 \hbar}}\frac{l}{i^l \, \tau}
J_{l}\left[\frac{V \tau}{2 \hbar}\right] d\tau \,,
\end{align}

we get the following estimate
\begin{subequations}
\begin{align}
&\left|u(l,t)\right|=\left|
\int_{t}^{\infty}e^{\frac{i p^2 \tau}{2m \hbar}-\frac{i V \tau}{2 \hbar}}\frac{l}{\tau}
J_{l}\left[\frac{V t}{2 \hbar}\right] d\tau
\right|\,\leq
\int_{\frac{t V}{2 \hbar}}^{\infty}
\left|\frac{l\,J_{l}(y)}{y}\right| dy\,\leq \\
&\left(\int_{\frac{t V}{2 \hbar}}^{\infty}\frac{l}{y^{1+2\epsilon}} dy \right)^{\frac{1}{2}}
\cdot
\left(\int_{0}^{\infty}
\frac{\left(J_{l}(y)\right)^2}{y^{1-2\epsilon}}
\right)^{\frac{1}{2}}\,=\,\\
& l \frac{1}{\sqrt{2 \epsilon}} \left(\frac{2 \hbar}{V t}\right)^{\epsilon}
\cdot
\left(
2^{2 \epsilon}
\frac{\Gamma [1-2 \epsilon]\Gamma[l+\epsilon]}
{2 (\Gamma[1-\epsilon])^2 \Gamma[1+l-\epsilon]}
\right)^{\frac{1}{2}} \,\leq \\
\label{est}
& l \frac{1}{\sqrt{2 \epsilon}}\left(\frac{2 \hbar}{V t}\right)^{\epsilon}
\cdot
\left(
2^{2 \epsilon}
\frac{\Gamma [1-2 \epsilon]}
{2 (\Gamma[1-\epsilon])^2 }
\right)^{\frac{1}{2}}
\quad \mbox{with} \quad 0\,<\,\epsilon\,<\,\frac{1}{2}\,,
\end{align}
\end{subequations}

where we have applied the Schwarz inequality and the integral formula
(\cite{Magnus})
\begin{equation}
\label{Bessel2}
\int_{0}^{\infty}
\frac{\left(J_{l}(y)\right)^2}{y^{1-2\epsilon}}=
2^{2 \epsilon}
\frac{\Gamma [1-2 \epsilon]\Gamma[l+\epsilon]}
{2 (\Gamma[1-\epsilon])^2 \Gamma[1+l-\epsilon]}\,.
\end{equation}

So we conclude setting $\epsilon=1/4$ for $\Delta_{l}$ (\ref{Delta})
\begin{equation}
\left|\Delta_{l}\right|\sim O(l)\cdot 
O\left(\left[\frac{2 \hbar}{V t}\right]^{\frac{1}{4}}\right)
\end{equation}

\section{Decreasing behaviour in momentum space of functions with compact support and the examination of the used
assumptions about orders of magnitude based on an example} 
\label{Consistency}
According to the Palay Wiener theorem (\cite{Za}), functions with compact support in position space can not be restricted 
to a finite interval in momentum space as well.
 Nevertheless the concentration of those functions in momentum space around their 
expectation value can be shown 
to be prescribed by the Fourier transform of a  generic  reference function. We choose an appropriate wavepacket with a reference 
function of Gaussian shape and derive the conditions that justify (\ref{Rand2},\ref{Approxp1},\ref{Faktor2}). We also show that we can use the reference function
for the evaluation of (\ref{PsiTunnel2}) as we did for the example presented in figure \ref{Blubber}.

We start with a normalized wavefuntion $\psi(x)$ with position expectation value $x_{0}$ that 
is assumed to be zero for $|x-x_{0}|\,\geq\,B$. Let $\psi_{0}(x)$ be a generic reference function that fulfills
\[
\begin{array}{lll}
\psi_{0}(x)=\psi(x)&\qquad \mbox{for} \qquad & |x-x_{0}|\,\leq\,B\\
\psi_{0}(x)=-\delta \psi(x) &\qquad  \mbox{for} \qquad & |x-x_{0}|\,>\,B\quad,
 \end{array}
\]  
where $\delta \psi$ is zero for $|x-x_{0}|\,\leq \,B$. Then $\psi(x)$ is represented by the sum
\begin{equation}
\label{Summe}
 \psi(x)=\psi_{0}(x)+\delta \psi (x)
\end{equation}

If the norm of $\delta \psi$ is given by $\epsilon$ we find for the reference function

\[
 \langle\psi_{0} |\psi_{0} \rangle\,=1+\epsilon^2\,.
\]

The representation of $\psi(x)$ in momentum space reads

\[
 f(p)=f_{0}(p)+\delta f(p) 
\]
where $f_{0}(p)$ and $\delta f(p)$ are the representations  $\psi_{0}(x)$ and $\delta\psi(x)$ in momentum space.

Using the Schwarz inequality and taking into account that the Fourier transform preserves the $L_{2}$ norm we 
get for the integral over the momentum density in the region outside the interval $(p_{min},p_{max})$

\begin{align}
\label{Rechnung}
&\int_{-\infty}^{p_{min}} f^{*}(p)f(p)\, dp + \int_{p_{max}}^{\infty} f^{*}(p)f(p)\, dp \,\leq \\
\nonumber
&\int_{-\infty}^{p_{min}} f_{0}^{*}(p)f_{0}(p)\,dp+ \int_{p_{max}}^{\infty} f_{0}^{*}(p)f_{0}(p) dp\,+\epsilon^2 +
2 \sqrt{1+\epsilon^2}\epsilon\,.
\end{align}

In order to provide an explicit example, we choose the initial wavepacket with the position and momentum
expectation values $x_{0},p_{0}$
\begin{subequations}
\label{initial}
\begin{align}
\label{initiala}
& \psi (x)=N^{-1/2} 
e^{-\frac{i p_{0} x}{\hbar}}
 e^{-\frac{(x - x_{0})^2} { 2 a}}
 \left[-L + \frac{x - x_{0}}{\sqrt{a}}\right]^2 
\left[L + \frac{x - x_{0}}{\sqrt{a}}\right]^2 \quad 
&\mbox{for}\quad  |x-x_{0}|\,<\,L \sqrt{a} 
 \\
 & \psi(x)=0 \quad 
 &\mbox{for}\quad |x-x_{0}|\,\geq\,L \sqrt{a}\,,
 \end{align}
 \end{subequations}
 where the normalization constant $N$  is given by
 \[
N= \frac{\sqrt{a}}{16} e^{-L^2 }(2 L (-105 + 50 L^2 - 20 L^4 + 8 L^6) + 
   e^{L^2} (105 + 8 L^2 (-15 + 9 L^2 - 4 L^4 + 2 L^6)) \sqrt{\pi}
    \mbox{Erf}[L])\,.
 \]
The factors multiplied to the Gaussian ensure that $\psi(x)$ is continuously differentiable at $x=x_{0} \pm L \sqrt{a}$.

The position and momentum uncertainty are determined by
\begin{subequations}
\label{Varianz}
\begin{align}
 &(\Delta x) ^{2}\, =
\frac{a (2 L (-945 + 210 L^2 - 52 L^4 + 8 L^6) + 
 e^{L^2} (945 - 840 L^2 + 360 L^4 - 96 L^6 + 16 L^8) \sqrt{\pi}
    \mbox{Erf}[L])}{
4 L (-105 + 50 L^2 - 20 L^4 + 8 L^6) + 
 2 e^{L^2} (105 - 120 L^2 + 72 L^4 - 32 L^6 + 16 L^8) \sqrt{\pi}
   \mbox{Erf}[L]} \\
&(\Delta p) ^{2}\,=
\frac{\hbar^2 2 L (-225 + 18 L^2 + 12 L^4 + 8 L^6) + 
 e^{L^2} (225 + 8 L^2 (-21 + 5 L^2 + 4 L^4 + 2 L^6)) \sqrt{\pi}
   \mbox{Erf}[L])}{2 a (2 L (-105 + 50 L^2 - 20 L^4 + 8 L^6) + 
   e^{L^2} (105 - 120 L^2 + 72 L^4 - 32 L^6 + 16 L^8) \sqrt{\pi}
     \mbox{Erf}[L])} \,,
\end{align}
\end{subequations}

where
\begin{align*}
  & \lim_{L \rightarrow \infty}(\Delta x) ^{2}\,=\frac{a}{2}\quad \mbox{and} \quad &\Delta x ^{2}\leq \frac{a}{2} \quad
  \mbox{for} \quad L\,>\,2
  \\
&\lim_{L \rightarrow \infty}(\Delta p) ^{2}\,=\frac{\hbar^2 }{2 a}
\quad \mbox{and} \quad &\Delta p ^{2}\leq 1.2 \frac{\hbar^2}{a}\quad
  \mbox{for} \quad L\,>\,2
\end{align*}

So we see that it is possible to choose the momentum uncertainty sufficiently small to justify (\ref{Approxp1}).
We take as  reference function $\psi_{0}(x)$ the extension of (\ref{initiala}) to the whole real line. We find for the norm of the difference
function $\delta{\psi}$ according to (\ref{Summe})
\begin{equation}
\epsilon^2 =\frac{
2 L (105 - 50 L^2 + 20 L^4 - 8 L^6) + 
 e^{L^2} (105 + 8 L^2 (-15 + 9 L^2 - 4 L^4 + 2 L^6)) \sqrt{\pi}
  \mbox{ Erfc}[L]
  }
  {
2 L (-105 + 50 L^2 - 20 L^4 + 8 L^6) + 
 e^{L^2} (105 + 8 L^2 (-15 + 9 L^2 - 4 L^4 + 2 L^6)) \sqrt{\pi}\mbox{Erf}[L]\,,
 }
\end{equation}

where                                                        
 \begin{equation}
  \label{epsilon}
 \lim_{L \rightarrow \infty}\epsilon\,=\sqrt{\frac{24}{\sqrt{\pi}}}e^{-L^2/2}L^{-9/2}
 \mbox{and} \quad  \epsilon \leq \sqrt{\frac{24}{\sqrt{\pi}}}e^{-L^2/2}L^{-9/2} \quad
  \mbox{for} \quad L\,>\,3
  \end{equation}
For the further calculations we introduce the dimensionless quantities
\begin{equation}
 \label{scaled}
 P=\frac{\sqrt{a}}{\hbar} p\,,\,\quad  P_{0}=\frac{\sqrt{a}}{\hbar} p_{0}\,,
 \quad X=\frac{x}{\sqrt{a}}\,,\quad  X_{0}=\frac{x_{0}}{\sqrt{a}}\,,\quad D=\frac{d}{\sqrt{a}}\,.
\end{equation}
We find for the momentum representation of the reference function $\psi_{0}(x)$ 
\begin{equation}
 \label{RMomentum}
 f(p)=\frac{a^{1/4}}{\sqrt{h}}\frac{4  e^{-(1/2) (P - P_{0}) (P - P_{0} + 2 I X_{0})} (3 + L^4 + 
   2 L^2 (-1 + (P - P_{0})^2) - 6 (P - P_{0})^2 + (P - P_{0})^4)}{\sqrt{
2 e^{-L^2}  L (-105 + 50 L^2 - 20 L^4 + 8 L^6) + 
  (105 + 8 L^2 (-15 + 9 L^2 - 4 L^4 + 2 L^6)) \sqrt{\pi} \mbox{Erf}[L]}}\,.
\end{equation}
Assuming that the interval $(p_{min},p_{max})$ is symmetric around $p_{0}\,>\,0$ we find
for the probability outside the interval
\begin{align}
 \label{PRand}
 &\mathcal{P}_{rest}\equiv \int_{-\infty}^{p_{min}} f_{0}^{*}(p)f_{0}(p)+
 \int_{p_{max}}^{\infty} f_{0}^{*}(p)f_{0}(p) dp =
 \\ \nonumber
& 
\left\{
 2 e^{-P_0^2(1+(K-2)K)} 
(-1 + K) P_0 \left(-39 + 72 L^2 - 88 L^4 + 32 L^6 +\right. \right.\\ \nonumber
  & 
 \left. 2 (-1 + K)^2 (83 + 24 L^2 (-3 + L^2)) P_0^2 + 
    4 (-1 + K)^4 (-17 + 8 L^2) P_0^4 + 8 (-1 + K)^6 P_0^6 \right) 
 \\ \nonumber
  & 
  +\left.
 \left(105 + 
      8 L^2 (-15 + 9 L^2 - 4 L^4 + 2 L^6)\right) \sqrt{\pi}
     \mbox{Erfc}[P_0 (K-1)]
  \right\}
 \cdot
\\ \nonumber
& \left(
 Abs\left[
2 e^{-L^2}  L (-105 + 50 L^2 - 20 L^4 + 8 L^6) + 
   (105 + 8 L^2 (-15 + 9 L^2 - 4 L^4 + 2 L^6)) \sqrt{\pi}\mbox{ Erf}[L]
 \right]
    \right)^{-1} \,,
 \end{align}
where $K \equiv p_{max}/p_{0}$.

According to (\ref{Rechnung}) the application of the approximation (\ref{Rand2}) means that we neglect a portion of the probability 
of the magnitude $\epsilon^2 + \mathcal{P}_{Rest}$. Since the wavepackets that leave the barrier are of the magnitude 
 \[
  e^{\frac{ -d(2l+1) \sqrt{2 m V-p_{0}^2}}{\hbar}}=e^{-(2l+1) D P_{0}\sqrt{\frac{1-k_0}{k_0}}}\,.
  \]
  our results are relevant compared to the neglected parts if the condition

  \begin{equation}
  \label{condition}
 D P_{0}\,(2l+1)
 \sqrt{\frac{1-k_0}{k_0}}
\lesssim \mbox{ln}[\epsilon+\mathcal{P}_{Rest}]   
  \end{equation}
  
  is fulfilled.

For the initial wavfunction evaluated in fig. \ref{Blubber} with the parameters 
\begin{equation}
\label{parameters}
 P_{0}=10,\,k_{0}=1/2,\,X_{0}=-20,\,L=20
 \end{equation}
and with the choice 
\begin{equation}
K=p_{max}/p_{0}=1.4\,,
\end{equation}
(\ref{condition}) reads
\begin{equation}
 \label{Bcondition}
 D (2l+1)\cdot 10\,\,\lesssim \, 18\,.
\end{equation}

So the wavepackets up to  $D (2l+1)\leq 1.8$ meet \ref{condition}. Moreover the set of parameters fulfills the requirements of (\ref{Faktor2}), since
\begin{equation}
\frac{1}{\sqrt{1-\frac{p_{max}^2}{2 m V}}}=
\frac{1}{\sqrt{1-\frac{p_{max}^2}{p_{0}^2}}}\approx 7\,.
\end{equation}

For the evaluation of the first three terms of (\ref{PsiTunnel2}) in fig.\ref{Blubber} we have used $f_{0}(p)$ instead of $f(p)$: This is justified since for our choice
of parameters (\ref{parameters})  $\epsilon \ll 1$ . Moreover the obtained position shift (\ref{Shift})
\[
\delta x_{l} = \frac{2 (1+2l) \hbar}{\sqrt{2 m V-p^2_{0}}}= 
2 (1+2l)\frac{\sqrt{a}}{P_{0}}
\sqrt{\frac{k_{0}}{1-k_{0}}}
\]
is bigger than the correction of the positions expectation value $x_{0}$ caused by $\delta \psi$. We find

\begin{align}
& \delta x_{ref}\equiv
 \langle\psi_{0}\, x\, \psi_{0} \rangle\ - \langle\psi\, x\, \psi \rangle\ = \langle\delta\psi\, x\,\delta \psi \rangle\ =
 \\  
& x_{0}\cdot\frac{2 L (105 - 50 L^2 + 20 L^4 - 8 L^6) + 
 e^{L^2} (105 + 8 L^2 (-15 + 9 L^2 - 4 L^4 + 2 L^6)) \sqrt{\pi}
   \mbox{Erfc}[L]}{
2 L (-105 + 50 L^2 - 20 L^4 + 8 L^6) + 
 e^{L^2} (105 + 8 L^2 (-15 + 9 L^2 - 4 L^4 + 2 L^6)) \sqrt{\pi} \mbox{Erf}[L]}\,.
\end{align}
Since 
\[
\left| \delta x_{ref}\right|\,\leq \left|x_{0}\right|\cdot \frac{24}{\sqrt{\pi}}e^{-L^2}L^{-9}\quad \mbox{for} \quad L\,>\,2\,,
\]
the condition $\delta x_{ref}\, \ll\, \delta_{l}$ is fulfilled if
\[
 \frac{24}{\sqrt{\pi}}e^{-L^2}L^{-9}\, 
 \ll \,2(1+2l) P_{0}^{-1}\left|X_{0}^{-1}\right| 
 \sqrt{\frac{k_{0}}{1-k_{0}}}\,,
\]

which is the case for the choice of parameters (\ref{parameters}).


\begin{thebibliography}{99}
\bibitem{Hauge}E.H.Hauge,J.A. Stovneng (1989):
Tunneling Times: A critical review, Rev.Mod.Phys. 61, 917
\bibitem{Landauer}  R.Landauer , Th.Martin (1994): Barrier interaction time in tunneling, Rev.Mod.Phys.66, 217
\bibitem{Winful} H.Winful (2006): Tunneling time, the Hartmann effect and superluminality:
 A proposed resolution of an old paradox, Phys. Rep.436,1-69
 \bibitem{Olkhovsky} V. Olkhovsky et. al. (2004): Unified analysis of photon
 and particle tunneling Phys. Rep. 398, 133-178
 \bibitem{Razavy} M.Razavy: Quantum theory of Tunneling, World Scientific Singapore 2003
  \bibitem{Galapon}E.Galapon (2012): Only above barrier energy components contribute
  to barrier traversal time, Phys.Rev.Lett.108,170402
 \bibitem{Landsmann} A. Landsman, U.Keller (2014: Tunneling time in strong field ionization, J.Phys.B 47, 204024
 \bibitem{Landsmann1} A. Landsman, U.Keller (2015: Attosecond science and the tunneling time problem, Phys.Rep.547,1-24
\bibitem{Wang} Z.S.Wang et.al.(2004: Quantum tunneling time, Phys.Rev. A 69, 052108
\bibitem{Cheng} Y. Cheng (2010): On the tunneling time of arbitrary continuous potentials and the Hartmann effect, Chin.Phys.B, 19,No.11, 117305
\bibitem{Pfeiffer} A.Pfeiffer et. at. (2012):
Attoclock reveals natural coordinates of the 
laser-induced tunneling current flow in atoms, Nature Physics Vol. 8, 76
\bibitem{Eckle} P.Eckle et.al. (2008): Science Vol.322 1525
\bibitem{Landsmann2} A.Landsman (2014): Ultrafast resolution
\bibitem{Hartmann} T. Hartmann(1962): Tunneling of a Wave Packet, J. Appl. Phys. 33, 3427
of tunneling delay time, Optica Vol.1,No.3,343
  \bibitem{MacColl} L.A.MacColl(1932): Note on the Transmission and Reflection of Wave Packets by Potential
                  Barriers, Phys.Rev. 40/ May 1932/621-626
 
 \bibitem{Hasan} M. Hasan, B.Mandal (2018): Phys.Lett. A382, 248-252
 \bibitem{Steinberg}A. Steinberg et. al. (1993): Measurement of single-photon tunneling time, Phys.Rev.Lett. 71/5,708
 \bibitem{Spielmann} Ch. Spielmann, R. Szip\"os, A.Stingl, F. Krausz (1994) :
Tunneling of optical pulses throug photonic band gaps, Phys.Rev. Lett. 73, 2308
\bibitem{Longhi} S.Longhi et. al (2001):Superluminal optical pulse propagation, Phys. Rev. E 64055602
\bibitem{Chiao} R.Y. Chiao et. al.(1993): Faster than light?, Scientific American 269, 52
\bibitem{Winful1} H.G. Winful (2003): Mechanism for 'superluminal' tunneling, Nature 424,638

 \bibitem{Methode} N.Riahi (2017): Solving the time-dependent Schr\"odinger
equation via Laplace transform, Quantum Stud.: Math. Found.  4/2, 103-126

\bibitem{Bernardini} A.E. Bernardini (2009): Stationary phase method and delay times for relativistic
 and non-relativistic tunneling particles, Ann.of Phys. 324, 1303-1339
\bibitem{Pereyra} P.Pereyra (2000): Closed formulas for tunneling time in superlattices,
Phys. Rev.Lett. 84/8, 1772
\bibitem{LosTunnel}
V.F.Los,M.V. Los (2012) : A time-dependent exact solution for wave-packet scattering at a rectangular barrier,
J.Phys.A: Math.Theor. 45,095302
 \bibitem{Bounce}
 M.A.Doncheski et. al. (1999): Anatomy of a quantum 'bounce',Eur.J.Phys.20, 29-37
 \bibitem{Erdely} A.Erdely: Tables of Integral Transforms, McGraw-Hill Book Company, 1954
 \bibitem{Magnus} W.Magnus, F.Oberhettinger, R.P.Soni: Formulas and Theorems for the 
 Special Functions of Mathematical Physics, Springer-Verlag 
  \bibitem{Za} A.Zayed: Handbook of Function and Generalized Function Transformations, CRC Press 1996 
 
 

\end{thebibliography}
\end{document}